\documentclass[twocolumn,showpacs,floatfix,prl]{revtex4}
\usepackage{amssymb}
\usepackage{graphicx}
\usepackage{dcolumn}
\usepackage{bm}
\usepackage[usenames]{color}

\begin{document}

\title{On the nature of the optical conductivity of polaron}
\author{G.~De~Filippis$^1$, V.~Cataudella$^1$, A.~S.~Mishchenko$^{2,3}$ and
N.~Nagaosa$^{2,4}$}
\affiliation{$^1$SPIN-CNR and Dip. di Scienze Fisiche - Universit\`{a} di Napoli
Federico II - I-80126 Napoli, Italy \\
$^2$ Cross-Correlated Materials Research Group (CMRG), ASI, RIKEN,
Wako 351-0198, Japan
\\
$^3$RRC ``Kurchatov Institute'' - 123182 - Moscow - Russia\\
$^4$Department of Applied Physics, The University of Tokyo,
7-3-1 Hongo, Bunkyo-ku, Tokyo 113, Japan}


\begin{abstract}
We study theoretically a fundamental issue in solids: 
the evolution of the optical spectra of polaron
as the electron-phonon coupling increases. By comparing the exact results
 obtained by diagrammatic Monte Carlo method and the data
got through exact diagonalization within an appropriate subspace 
of the phononic wavefunctions, 
the physical nature
 of the crossover from weak to strong coupling is revealed. 
The optical spectra are well understood by the quantum mechanical 
superposition of states with light and heavy phonon clouds 
corresponding to large and small polarons, respectively. 
It is also found that the strong coupling Franck-Condon regime is not
 accessible at realistic values of the coupling constant.
\end{abstract}

\maketitle

Polaron physics plays a crucial role in describing
magnetoresistive perovskites \cite{millis}, high 
temperature superconductors \cite{Shen,MisO}, molecular 
semiconductors \cite{troisi}, fullerenes \cite{han}, polar semiconductors \cite{dev}, etc.
However, in spite of a lot of efforts, 
there are still only a few well grounded results 
concerning spectral properties,
optical conductivity (OC) and Lehman spectral function (LSF) of polaronic systems at any coupling. 
Even the simplest model for periodic system, i.e. Holstein model, 
is not an exception and the nature of its spectral response at 
all couplings is still a puzzle, in particular in nearly adiabatic case, 
the most frequently encountered regime in experiments.

In contrast to Frochlich-like models \cite{Frohlich},  
which simplifies the manifold of all momenta of the lattice 
system by using continuum approximation, 
the Holstein model takes into account the true
momentum space of the system \cite{Holstein}. 
Here the tight-binded charge carrier is coupled with an optical local phonon mode:  
\begin{eqnarray}
H=-t \sum_{<i,j>} c^{\dagger}_{i}c_{j} +\omega_0
\sum_{i}b_{i}^{\dagger} b_{i}+ g \omega_0 \sum_{i} c_{i}^{\dagger
}c_{i} \left( b_{i}^{\dagger} +b_{i} \right) \nonumber \; ,
\end{eqnarray}
where $t$ is the nearest neighbor hopping amplitude, 
$c_{i}^{\dagger }$ ($c_{i}$) denotes electron creation (annihilation) operator 
in the site $i$, and $b_{i}^{\dagger }$ ($b_{i})$ creates
(annihilates) a phonon in the site $i$ with frequency $\omega_0$. 
The strength of electron-phonon interaction (EPI) is characterized 
by dimensionless coupling parameter $\lambda=g^2 \omega_0/2tD$
expressed in terms of coupling constant $g$ and system 
dimensionality $D$.

There are many methods able to calculate the ground state 
(GS) properties of the Holstein polaron, see 
Ref.[\onlinecite{books}] as an incomplete list. 
In particular adiabatic and some variational calculations lead to discontinous 
transition between weak and strong coupling regimes as EPI increases \cite{toyozawa}. 
The quantum correction to this picture is a singular perturbation, and essentially non perturbative. 
A naive picture is that the discontinous transition becomes a crossover by quantum mixture of states 
describing asymptotic regimes. 
On the other hand, the situation changes drastically if some spectral properties, e.g. OC, are considered.  
The main problem is due to the infinite dimensional phononic 
Hilbert space, that has prevented an in-depth analysis so far. 
The full phonon basis exact diagonalization (ED) method 
\cite{EDS} is basically valid only for 1D and 2D systems 
and suffers from effects caused by very small 
size of the considered clusters, normally 
limited to 10 sites.
The dynamical mean-field theory
 \cite{dmft}, though it is exact in infinite dimension system, leaves 
doubts about applicability to realistic 3D, 2D, and especially 1D
systems. 
Finally, containing no approximation, Diagrammatic Monte Carlo
 (DMC) methods \cite{DMC_OC,MisO} provide approximation free 
results for infinite systems of any dimension. 
However, none of the above methods helps physical intuition or provides an 
understanding of the nature of OC 
and/or excited states in the weak, strong and especially 
in traditionally most puzzling intermediate coupling regime. 
An exception is momentum average (MA) \cite{MA2} approach.   
It suggests an analytic formula and explicitly relates 
the shape of OC in the strong coupling regime to the 
phononic cloud dragged 
by polaron moving through the crystal.
Unfortunately, the intermediate coupling regime, the most interesting one from an experimental point of view, 
is a puzzle for the MA approach too. 

In the present letter we present the exact DMC results for the 
OC of 1D system for any interaction value ranging from 
the weak- to the strong-coupling regime. 
The regular part of the OC, a typical experimental probe of a physical system, is defined \cite{shastry} as:
\begin{eqnarray}
\sigma_{reg}(\omega)= \pi \sum_{n \ne 0} \frac{
| \left\langle \psi_n \right | j
\left | \psi_0 \right\rangle |^2} {\omega_n} \left [
\delta \left( \omega-\omega_n \right ) + \delta \left( \omega+\omega_n \right )
 \right ]  \nonumber \; .
\label{sigmareg}
\end{eqnarray}
Here
\begin{equation}
j=i e t \sum_{<i,j>} c^{\dagger}_{i}c_{j}
\left ( \vec{R_i}- \vec{R_j}\right )
\label{cur}
\end{equation}
is the current operator, $\vec{R_i}$ is the position vector of the site
$i$, $e$ is charge, $ \left | \psi_n  \right\rangle$ and $E_n$ are the
exact eigenstates and eigenvalues, and $\omega_n=(E_n-E_0)$.
The lattice parameter $a$ and Planck constant
 $\hslash$ are set to unity.

\begin{figure}
        \includegraphics[scale=0.64]{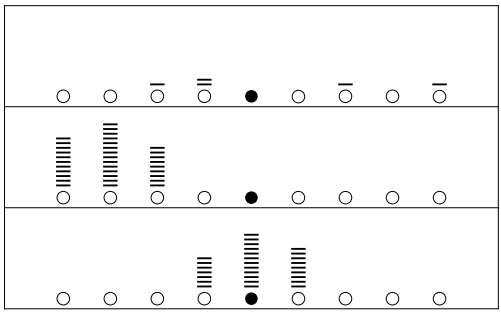}
        \caption{The empty (full) circles stand for the lattice sites (electron); the dashes denote phonons.
The upper panel represents a typical WCPC and the two
lowest panels show possible realizations of the SCPC.}
\label{figbasis}
\end{figure}

\begin{figure}[b]
\flushleft
        \includegraphics[scale=0.425]{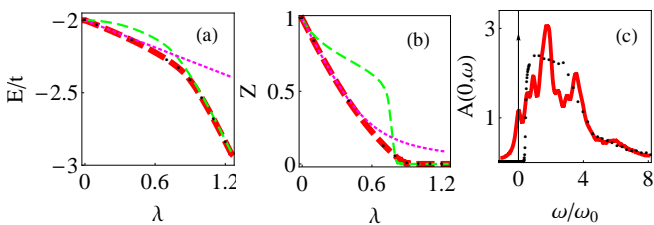}
        \caption{(Color online) (a) Energy versus $\lambda$ in
the DMC (black dots), DPC (dashed thick red line),
lowest order perturbation \cite{Mahan} (dotted magenta line),
and FC (dashed thin green line) approaches; (b)
GS spectral weight versus $\lambda$ in the DMC (black dots),
DPC (dashed thick red line), WCPC (dotted magenta line),
and MA(0) approaches (dashed thin green line).
(c) Lehman spectral function
in the DMC method (black dots for incoherent part and
vertical black arrow for $\delta$-functional polaron state)
and in DPC approach (solid red line).}
\label{fig1}
\end{figure}

To disclose the nature of the OC 
we introduce two contrasting physical pictures of polaronic phonon cloud: 
i) weak coupling phonon cloud (WCPC), that allows a very good description
of the weak-coupling regime; ii) strong coupling phonon 
cloud (SCPC), that is able to reproduce 
accurately the properties of the strong-coupling regime. 
The surprising result obtained by our analysis of the exact DMC data  
is that a straightforward linear superposition of the 
weak- and strong-coupling phonon clouds, i.e. double phonon 
cloud (DPC), is capable of describing polaron 
properties in any coupling regime.
Comparison of the exact results with those obtained by the 
DPC analysis points out that 
the Franck-Condon (FC) approach, based on the factorization of the 
wavefunctions associated to charge and lattice degrees of freedom and on optical transitions 
in a frozen lattice, fails at all couplings except for strong and unrealistic 
values of charge-lattice interaction, i.e. at couplings where one can use the 
crudest photoemission approximation (PA) that 
describes the photoionization process towards the free-electron
continuum: within the FC approach one replaces
the exact excited charge states within the GS lattice potential well with the free electron states
in the absence of EPI. 

\begin{figure}
\flushleft
        \includegraphics[scale=0.42]{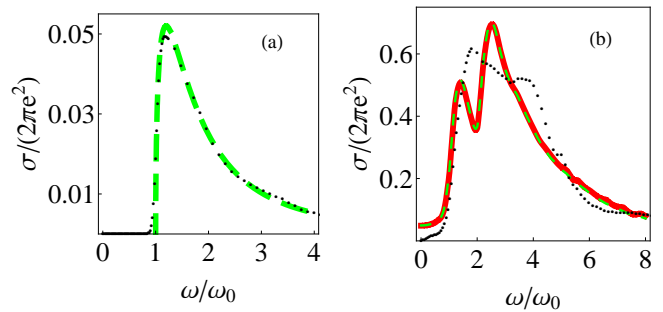}
        \caption{(Color online) OC in DMC method  (black dots)
at $\lambda=0.01$ (a) and $\lambda=0.4$ (b).
Dashed green line is lowest perturbative result \cite{Mahan} in (a).
Dashed thin green (solid thick red) line in (b) is the result of
WCPC (DPC) approach.}
\label{fig2}
\end{figure}

The phonon cloud in the weak coupling regime is not 
characterized by strong local deformations, and, thus, we set that the 
WCPC consists of up to $n \le {\cal N}_{wc}$ phonons situated at
arbitrary sites (upper panel in Fig. \ref{figbasis}).
To the contrary, the lattice deformation is 
well localized in the strong coupling regime and, thus, we 
construct the phonon SCPC as $n \le {\cal N}_{sc}$ phonons
located at 3 nearest neighbor sites.
Due to nonlocal nature of the current operator Eq. (\ref{cur}) it is 
crucial that the deformed neighboring sites are situated in 
the arbitrary place with respect to the charge carrier 
(two lower panels in Fig. \ref{figbasis}).  
We found that ${\cal N}_{wc}=5$ and ${\cal N}_{sc}=50$ is enough
to describe successfully the system in the 
range of the $\lambda$ values considered in the following and concerning  
the 1D case with $\omega_0 / t = 0.1$ (tha lattice size is $80$).

First, we compare the GS properties obtained by 
DPC model and those coming from the exact
DMC method and demonstrate a perfect agreement    
(see Fig. \ref{fig1}a-b). 
A very good agreement is seen even in the intermediate coupling regime 
suggesting that the DPC combination of the weak- and strong-coupling phonon clouds 
provides an accurate description 
in the whole range of coupling parameters. 
Indeed, there are domains of $\lambda$ values where both strong-
and weak-coupling phonon cloud models fail but, nevertheless, 
the results using DPC model are still in perfect 
agreement with exact ones.  
Not only GS properties but also the LSF is well
reproduced by the DPC picture too (Fig. \ref{fig1}c). 
One can see the obvious failure of the simplest variant of the 
MA approach, (i.e. MA(0) that takes into account only lattice deformations 
located at one lattice site at a time), in describing the GS spectral weight $Z$ 
(see Fig. \ref{fig1}b). 
This failure explicitly demonstrates that 
in the intermediate coupling regime the phonon wavefunctions
are characterized by lattice distortions involving  
many sites at the same time.

Good agreement of the OCs got by perturbative and DMC 
approaches at $\lambda=0.01$ (see Fig. \ref{fig2}a) indicates 
that the final states of the optical transition do not contain more than one phonon.
To the contrary, evident rise of OC at $2\omega_0$ for $\lambda=0.4$ (see Fig. \ref{fig2}b) manifests 
importance of the processes where 
correlation between two successively emitted phonons is present.    
The agreement between WCPC approximation and DPC approach indicates that 
the contribution of the strong coupling counterpart of the
phonon cloud is still negligible at $\lambda=0.4$. 
 
\begin{figure}[t]
\flushleft
        \includegraphics[scale=0.42]{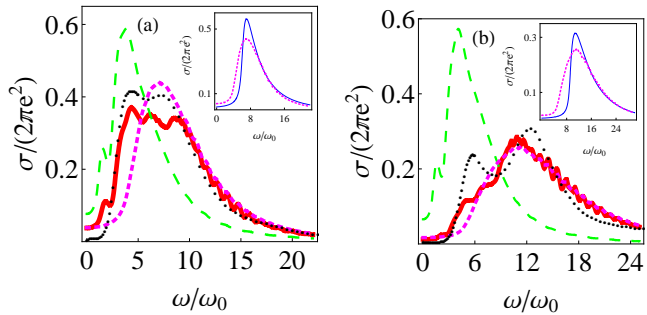}
        \caption{(Color online) OC in DMC (black dots),  DPC (solid  
red thick line),  WCPC (dashed green line),  and SCPC 
(dotted magenta line) approaches at $\lambda=0.77$ (a) and $\lambda=0.87$ (b). 
Insets show comparison of OCs in SCPC and FC (solid blue line)
approaches.
The OCs in the FC approach is shifted down by energy
$\approx 5 \omega_0$.}
\label{fig3}
\end{figure}

\begin{figure}[t]
\flushleft
        \includegraphics[scale=0.425]{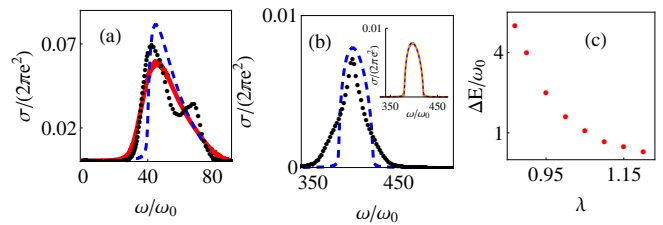}
        \caption{(Color online) OC in DMC (black dots), DCP
(solid red thick line), and FC (dashed blue line) approaches at $\lambda=1.5$ (a) and $\lambda=10$ (b).
Inset in (b) shows OC in FC approach and photoemission curve
(solid orange line).
Panel (b) shows the energy of lattice relaxation from the frozen
FC state to relaxed excited configuration of the SCPC state.}
\label{fig4}
\end{figure}

In Fig. \ref{fig3} the exact data for the OC in the 
intermediate coupling regime ($\lambda \approx 0.8$)
are compared with the results following from the WCPC, SCPC, 
and DPC conceptions of the phonon clouds. 
The intermediate coupling regime is a stumbling stone for most of 
the methods because any approach, starting from the characteristic 
features of either weak- or strong-coupling regime, fails there. 
The results in Fig. \ref{fig3} clearly reveal the reason for 
such difficulties, showing that the phonon cloud in this regime is a 
{\it superposition} of the WCPC and SCPC contributions. 
The dominating hump of the OC splits into two peaks: the former one 
is located at the energy determined by the weak coupling
 counterpart of the lattice deformation and the higher energy structure is governed by strong-coupling 
component of the phonon cloud.
Indeed we note that if the OC obtained in the framework of
the FC concept is shifted towards lower frequencies we 
recover the peak of the OC governed by strong-coupling lattice 
deformation (see insets in Fig. \ref{fig3}). 
Hence, the OC in the intermediate coupling regime is not yet
described by the FC processes where optical excitations correspond
to electronic excitations in the rigid lattice. 
The highest energy peak of the OC in the intermediate 
coupling regime can be interpreted as the result of light absorption 
in two successive steps: the lattice is frozen during the 
electronic transition at the first stage and then it relaxes 
and adapts itself to new electron configuration.
We studied the the energy of lattice relaxation from the frozen 
FC state $\Delta E$ and found that it becomes 
negligibly small only in the 
very strong coupling regime at $\lambda>1.2$ (Fig. \ref{fig4}c). 
The polaron mass at such couplings is three orders of magnitude larger than the bare mass.
Such effective mass, to our best knowledge, was never observed 
in any physical system. 

By increasing $\lambda$ there is a net transfer of spectral weight towards 
higher frequencies, and at $\lambda = 1.5$ only the 
higher energy contribution survives (see Fig. \ref{fig4}a). 
In this regime the energy of lattice relaxation from the frozen 
FC state is almost zero and the energies of the main peak of the OC 
in the DMC, DPC, SCPC, and FC approaches are the same.  
On the other hand there are differences in the tails of OC calculated 
in the above techniques, indicating the residual role played by the
nonadiabatic transitions even in the strong-coupling regime. 
Finally, the crudest PA approximation is valid at extremely large 
couplings (see inset in Fig. \ref{fig4}b).

Above discussed considerable success of the concept of the 
double phonon cloud paves a way to a novel all-coupling method called in the following 
Double Phonon Cloud ED (DPCED) approach. 
Limited phonon basis, including not the whole manifold of the phonon 
states but only the essential ones, enables studies of large 
systems up to about 100 sites. The approach is     
described in detail in the following. 

First of all we take into account the
translational invariance of the Hamiltonian and perform an
exact diagonalization of the Hamiltonian 
based on Lanczos algorithm, requiring that
the states have a definite momentum \cite{dagotto}.
Each of basis vectors is a linear superposition with
appropriate phases of the translational copies
(charge carrier and lattice configurations are
together rigidly translated) of a state having
the electron fixed at a site and phonon quanta located around it.

{\it Weak coupling counterpart} is restricted to the most 
$5$ phonons situated up to $5$ different sites which, in 
turn, are arbitrary
located with respect to the charge carrier.  
The phononic Hilbert subspace corresponding to the WCPC is generated by the following elements: 
\begin{equation}
\left | ph \right\rangle^{(WCB)}_{\left [ j \right],\left [ n_j \right]}  =
\prod_{h=1}^5
\frac { (a_{j_h}^{\dagger})^{n_{j_h}} \left | 0 \right\rangle_{j_h} } 
 {\sqrt{n_{j_h}!}} 
\prod_{i \ne \left[ j \right]}\left| 0\right\rangle_{i}.
\label{WCB}
\end{equation}
Here $i,j_h$ label the lattice sites, $\left [ j \right]=\left(j_h,h=1, ...5 \right)$ represents a set of 
$5$ different sites, 
$\left | 0 \right\rangle_i$ is the i-site phonon vacuum state,
and the integers $n_{j_h}=0, ....5$ are such that 
$\sum_{h} n_{j_h} \le 5$. 
In any basis element there are up to $5$ phonons distributed 
at the most 
on $5$ different sites: all the others sites are not deformed.  
In this way the scattering processes between the charge carrier and lattice up to $5$ phonons in $q$-space 
are exactly treated. 
This basis is able to recover the self-consistent Born 
approximation and goes beyond it including vertex corrections.

{\it Strong coupling counterpart} in the direct space representation 
contains the following set of states:
\begin{equation}
\left | ph \right\rangle^{(SCB)}_{j,n_j,n_{j'}}  =
\frac { (a_{j}^{\dagger})^{n_{j}} \left | 0 \right\rangle_{j} }
 {\sqrt{n_{j}!}} 
\prod_{j'} 
\frac { (a_{j'}^{\dagger})^{n_{j'}} \left | 0 \right\rangle_{j'} }
 {\sqrt{n_{j'}!}}
\prod_{i \ne \left(j,j' \right)}\left| 0\right\rangle_{i}.
\label{SCB}
\end{equation}
Here $j'$ indicates the nearest neighbors of $j$ site, 
and $n_{j'}$ represents the number of phonons on such sites. 
In any basis element there is a cluster with at the most 
$3$ deformed nearest neighbors sites which are located
at an arbitrary position with respect to the electron. 
We found that the sufficient basis for all studied above couplings 
is limited to $(n_j+\sum_{j'} n_{j'}) \le 50$.
We note that by restricting the cluster with excited 
phonons to one site 
one comes to the simplest formulation of the MA approximation, 
i.e. MA(0) \cite{MA}.   

{\it DPCED approach} is based on the idea that 
the physical properties for any value of the EPI can be described 
by diagonalizing the Hamiltonian in the phononic Hilbert
subspace generated by both above introduced
sets of states (excluding obvious double counting).

In conclusion, we have shown that 
the ground state, Lehman spectral function, and OC 
of the polaron at any coupling are 
well described in terms of superposition of two phonon clouds: the former one is 
spatially extended and contains 
small number of phonons, whereas the latter one is 
well localized and includes a large number of vibrational quanta.
Surprisingly, the lattice deformation around polaron in the 
puzzling intermediate coupling regime appears to be not a 
peculiar characteristic of this particular regime but a mere superposition 
 of quantum states representing two asymptotic couplings.
The accuracy of the above concept of double phonon cloud 
paves the way to a novel DPCED method, which uses not the whole manifold of the phonon 
states but only the essential ones, and enables study of 
large systems which are inaccessible by classic exact
diagonalization method. 
Comparison with the results for OC of 
the novel method with those 
obtained by approximation free 
Diagrammatic Monte Carlo method validates 
both the suggested double phonon cloud
concept and the novel numeric method based on it.   
The novel method can 
be further applied to problems of nonequilibrium dynamics 
\cite{han1} where such methods as diagrammatic Monte Carlo fail.   

ASM acknowledges support of RFBR 10-02-00047a.
NN is supported by MEXT Grand-in-Aid No.20740167, 19048008, 
19048015, and 21244053, 
Strategic International Cooperative Program (Joint Research Type) from 
Japan Science and Technology Agency, and by the Japan Society for the 
Promotion of Science (JSPS) through its ``Funding Program for
World-Leading Innovative R \& D on Science and Technology (FIRST Program)''.

\end{document}